\documentclass[conference]{IEEEtran}
\usepackage{graphicx}

\begin{document}
\title{Wireless Health Monitoring using Passive WiFi Sensing}

\author{\IEEEauthorblockN{Usman Mahmood Khan,
Zain Kabir,
Syed Ali Hassan}
\IEEEauthorblockA{School of Electrical Engineering \& Computer Science (SEECS), National University of Sciences \& Technology\\
(NUST), Islamabad, Pakistan \{13beeukhan, 13beezkabir, ali.hassan\}@seecs.nust.edu.pk}
}

\maketitle

\begin{abstract}
\noindent This paper presents a two-dimensional phase extraction system using passive WiFi sensing to monitor three basic elderly care activities including breathing rate, essential tremor and falls. Specifically, a WiFi signal is acquired through two channels where the first channel is the reference one, whereas the other signal is acquired by a passive receiver after reflection from the human target. Using signal processing of cross-ambiguity function, various features in the signal are extracted. The entire implementations are performed using software defined radios having directional antennas. We report the accuracy of our system in different conditions and environments and show that breathing rate can be measured with an accuracy of 87\% when there are no obstacles. We also show a 98\% accuracy in detecting falls and 93\% accuracy in classifying tremor. The results indicate that passive WiFi systems show great promise in replacing typical invasive health devices as standard tools for health care.

\end{abstract}

\begin{IEEEkeywords}
Passive WiFi sensing; breathing rate measurement; essential tremor classification; fall detection; phase extraction; Doppler; SDRs
\end{IEEEkeywords}

\IEEEpeerreviewmaketitle

\section{Introduction}
\noindent 

\noindent Over the past few years, there has been a growing interest in ubiquitous health monitoring [1-3]. One of the areas which benefited most from this surge in the interest is elderly care [4][5]. Today, we see wearable devices that continuously monitor vital health signs [6], track essential tremor in Parkinson patients [7], generate alarms if there are any falls, and do more. However, such devices present several challenges: they intrude with users' routine activities, they have to be worn all the time even during sleep, and they have to be frequently recharged. With the advances in wireless sensing, it has become possible to track and localize human motion [8-10]. In this paper, we explore if we can tap onto these advances to monitor three basic elderly care activities; i.e., breathing, tremor and falls. 

Breathing involves continuous inhale and exhale movements and may be used to study the subject's physiological state, her stress levels, or even emotions like \textit{fear} and \textit{relief}. Tremor is a movement disorder in which the subject experiences rhythmic shaking of a body part such as hands [7]. It is highly prevalent among older people and although not life threatening, it causes great inconvenience in social and daily life settings, such as writing and eating. Falls are also highly common among the ageing population and detecting them early is integral to effective interventions and subsequent treatments. 

Despite the ubiquitous nature of wireless networks, a little work has been done on exploring their suitability for health applications. This is concerning since the typical health monitoring techniques are inconvenient; they require contact with human body, and most of them are intrusive. For example, current breath monitoring solutions require a chest band [6], nasal probe [11], or pulse oximeter [12]. Technologies that are more comfortable such as wrist bands tend to be erroneous and unreliable. Tremor and fall monitoring devices are similarly inconvenient. However, the bigger concern is that contact devices are not suitable for elderly health care. Using a technology round the clock may be cumbersome or even demeaning for the elderly. Worse, they may be in a condition such as dementia where they can't remember to put on the device once they have worn it off.

Related literature on target detection and localization focuses on both active and passive radar techniques. Active radars employ dedicated transmitters, use high bandwidth and often require complex antenna arrays to go along with them [8]. In contrast, passive radars utilize the transmitted signals in the environment such as WiFi and cellular signals for target monitoring. Passive radars offer numerous advantages when compared to active radars, i.e., they are low-cost and covert due to their receive-only nature, they operate license free due to no bandwidth requirements, and they offer better Doppler resolution due to the possibility of higher integrations times. 

To the best of the authors' knowledge, this is the first comprehensive study on elderly-focused health care applications using passive WiFi sensing. Specifically, this research introduces a vital health wireless device (Wi-Vi) and makes the following contributions: 

\begin{itemize}
\item Proposes a phase extraction system in 2-D to determine breathing rate, classify essential tremor and detect falls.

\item Carries out extensive experiments to study the accuracy of elderly care applications in different environments and conditions.
\end{itemize}

\subsection{Related Work}
Multiple approaches have been proposed for human motion detection and other applications using passive sensing. Kotaru \textit{et.al} presented an indoor localization system using the channel state information (CSI) and received signal strength information (RSSI) [9]. Similar approaches have been used for other applications such as keystroke identification [13], in-home activity analysis [10], and virtual drawing [14]. Some work has been done on using active radars for breathing rate measurements. WiZ uses frequency modulated continuous wave (FMCW) signal chirps to detect breathing rate in both line-of-sight (LOS) and non line-of-sight (NLOS) scenarios [8]. Other similar works include an ultra wide-band radar with short transmit pulses to avoid reflections from surrounding objects [15], a continuous wave (CW) radar for clinical measurements [16], and a CW radar for indoor environment [17]. Essential tremor monitoring is mostly unexplored in the wireless domain and is typically done using electromyography sensors, accelerometers and gyroscopes [18]. 

The rest of the paper is organized as follows. In section 2, the phase extraction theory is discussed followed by an explanation of the 2-D system model and subsequent signal processing in section 3. Section 4 elaborates on the implementation procedure and section 5 lists the key results and limitations of this research. 

\section{Theory of Operation}

\noindent The observed frequency shift due to relative motion of transmitter and receiver is given by the well-know phenomenon of Doppler frequency, and is written as
\begin{equation}
{f_d = f_o \frac {v}{c} cos(\theta)},
\end{equation}

\noindent where \(f_o\) is the carrier frequency of the transmitted signal,  \(v\) is the speed of relative motion, \(c\) is the speed of light, and \(\theta\) is the angle of relative motion between transmitter and receiver.

\begin{figure}[!t]
\centering
\includegraphics[scale=0.48]{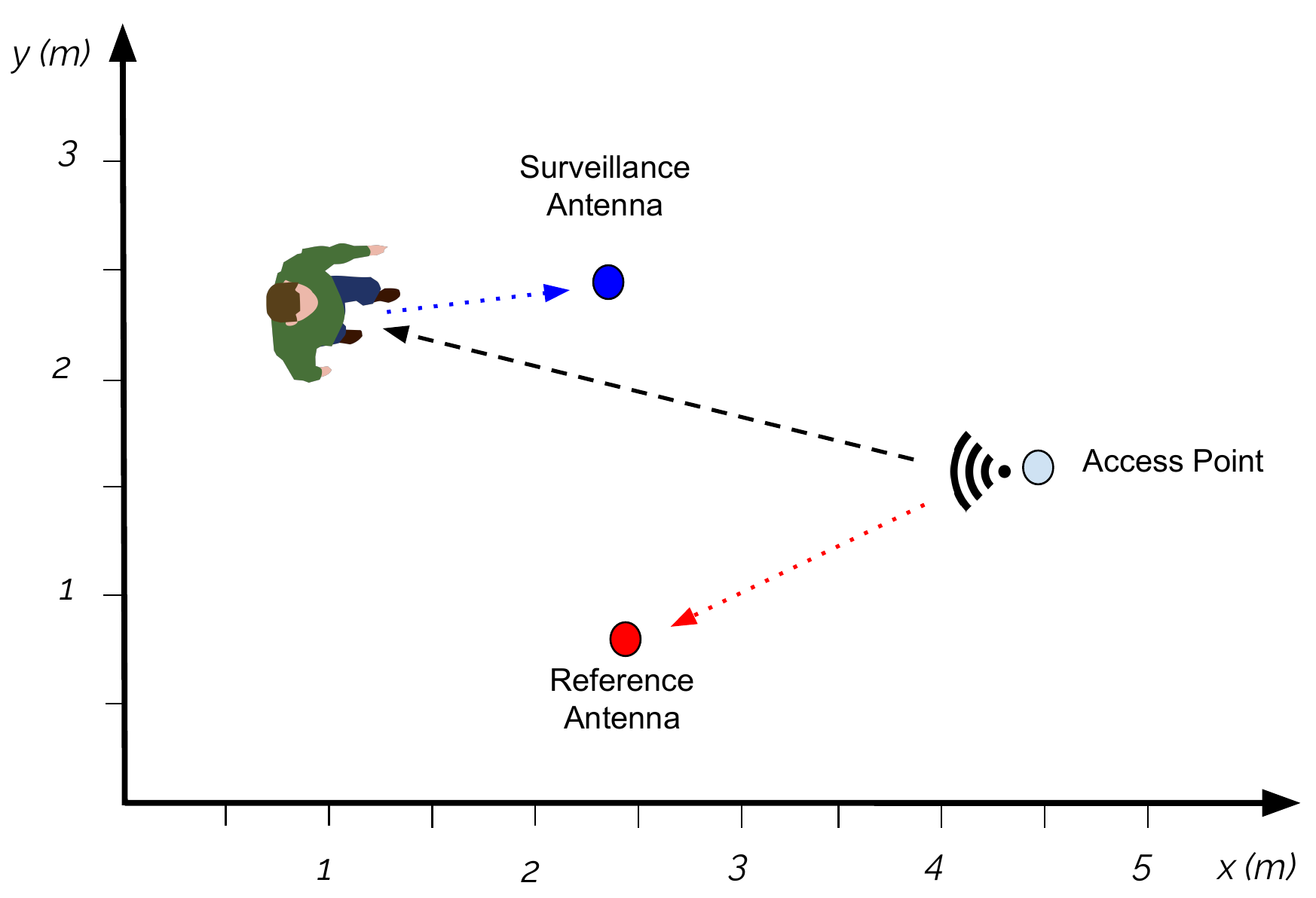}
\caption{Essential components of a passive radar system. Both surveillance and reference antennas pick up signals from the access point, but the surveillance signal contains an added Doppler shift due to human movement.}
\end{figure}

\noindent In a passive radar sensing system (shown in Fig. 1), one or multiple WiFi access points (APs) are used as transmitters, and spatially directional high gain antennas with narrow beam-widths are used to obtain the reference and the surveillance signals. One approach to obtain target range and Doppler information is by using the cross ambiguity function (CAF), which can be represented in its discrete form as
\begin{equation}
{\chi[\tau, f] = \sum_{n=0}^{N} r[n] s^{*}[n+\tau]e^{-i2\pi f_d\frac{n}{N}}},
\end{equation}

\noindent where \(r[n]\) is the reference signal, \(s[n]\) is the surveillance signal, \(\tau\) is the path delay from access point to the surveillance receiver, and \(N\) is the total number of samples in a single window. For the system given in (2), the total integration time is a function of total number of samples \(N\) and the sampling frequency \(f_s\). Accordingly, the Doppler resolution of the system is given as
\begin{equation}
{d = \frac{f_s}{N}} .
\end{equation}

\noindent Equation (3) shows that a long integration time is required to achieve a fine Doppler resolution. For example, an integration time of 5 seconds yields a Doppler resolution of 0.2Hz. This poses two challenges to \textit{Vi-Wi's} target applications which involve small scale body motions. First, due to the nature of these applications, multiple Doppler shifts will emerge in a single integration cycle resulting in an inherent ambiguity; second, positive and negative Doppler shifts may be present in a single cycle and thus cancel out the effects of one another.

\noindent In order to address these issues, Vi-Wi's operations consist of the following three steps:

\begin{itemize}
\item Extract phase information off the reflected surveillance signals through cross-correlation.

\item Develop a 2-D system model to identify phase variations in two separate planes.

\item Analyze and process phase variations to determine breathing rate, classify tremor and detect human falls.

\end{itemize}
In what follows, we present an elaborate approach to implement these steps.

\subsection{Phase Extraction}

\noindent Let a small scale movement in a single plane is given by \(m\). If the wavelength of transmitted signal is given by \(\lambda\), then the phase \(\phi\) associated with this movement can by described as

\begin{equation}
{\phi = \frac{2\pi m}{\lambda}}.
\end{equation}

\noindent Since the reference and the surveillance antennas are spatially separate, the small scale human motion is observed by surveillance antenna only. It follows then that the surveillance signal has an associated phase shift which is positive when the movement is towards the antenna and negative when the movement is away from it. Assuming that there is a single reflection at the surveillance antenna from the desired small scale motion, the surveillance signal is given as

\begin{equation}
{s[n] = Ax_{source}[n+\tau]e^{i2\pi f_{d}n}},
\end{equation}

\noindent where \(x_{source}[n]\) is the source signal, \(f_d\) is the Doppler shift, and \(A\) is the signal amplitude. In the simplest case when there is no movement, the Doppler shift is zero and the above equation simplifies to a delayed version of transmitted source signal with no phase shift. When there is a movement towards the plane of surveillance signal, the Doppler shift increases, resulting in a positive phase shift, and vice versa. Hence, by tracking phase variations in the reflected signal, we can also track the movements in the environment.

If we divide the data into \(B\) batches and assume that the time delay between reference and the surveillance signals is \(\tau\), then the cross correlation result in time domain for each batch \(b\), where \(b \in \{1,2,...,B\}\), is represented as
\begin{equation}
{y[b] = \sum_{n=1}^{N_b} r[i_b + n - \tau] s^{*}[i_b + n]},
\end{equation}

\noindent where \(N_b\) is the number of samples in each batch, and \(i_b\) is the starting sample in each batch. In (6), \((.)^{*}\) denotes the conjugate operator on a complex number. We note that since passive radars have a limited range resolution, therefore, \(\tau\) can be set to the maximum time delay our system is likely to encounter. If the sampling frequency is \(f_s\) and the time delay between the reference and surveillance signals is \(t_{lag}\), then the delay is given as
\begin{equation}
{\tau = \frac{f_s}{t_{lag}}}.
\end{equation}

\noindent In typical small scale motions such as breathing, the time delay is in the order of nano seconds (ns). Hence, we can assume \(\tau\) to be zero in context of our applications. In order to get smoother transitions in \(y[b]\) (and consequently in \(\phi[b]\)), we consider an \(x\%\) overlap between consecutive batches. The exact choice of \(x\) is governed by system's latency requirements. The phase of each batch, \(\phi[b]\), can now be found by taking inverse tangent of real and imaginary parts of complex series \(y[b]\), i.e.,
\begin{equation}
{\phi[b] = tan ^{ - 1}\frac{\Re(y[b])}{\Im(y[b])}}.
\end{equation}

\noindent Although \(\phi[b]\) encodes small scale body motions, it also captures reflections off static objects such as furniture and walls. In order to remove these time invariant phase shifts, we normalize \(\phi[b]\) to zero mean and unit variance, i.e.,
\begin{equation}
{\phi[b] = \frac{\phi[b] - \mu_b} {\sqrt{\frac{1}{W}\sum_{i=b-W+1}^{B} (\phi[i] - \mu_b)^2}}},
\end{equation}

\noindent where \(W\) denotes the window length, and \(\mu_b\) is given as
\begin{equation}
{\mu_b = \frac{1}{W}\sum_{i=b-W+1}^{B} \phi[i]}.
\end{equation}

\section{Passive Radar System Model in  2-D}

\begin{figure}[!t]
\centering
\includegraphics[scale=0.3]{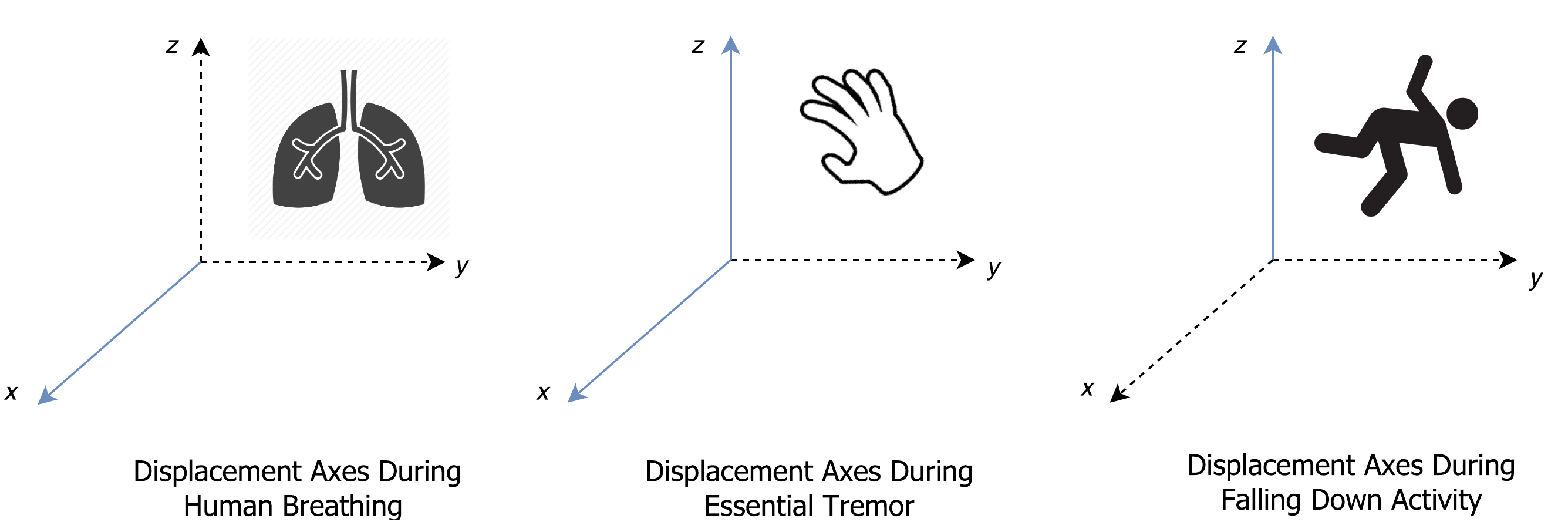}
\caption{Displacement axes during vital health care activities. Dotted lines show inactive axes for a specific activity.}
\end{figure}

\noindent Figure (2) shows the displacements involved in the  three activities of human breathing, essential tremor and human fall, respectively. Human respiration involves movements along a single axis which can be quantified by placing a surveillance antenna perpendicular to human chest. In contrast, essential tremor involves motion in two different planes which requires a minimum of two antennas placed perpendicular to one another. Finally, the motion involved in human fall is also along a single axis but perpendicular to the breathing motion.

\noindent In order to detect phase variations in these applications, Vi-Wi uses two surveillance antennas \(surv_1\) and \(surv_2\) with the following polar coordinate geometry:
\begin{equation}
{surv_1: \phi = \frac {-\pi}{2}, \theta = 0},
\end{equation}
\begin{equation}
{surv_2: \phi = \frac {-\pi}{2}, \theta = \frac {-\pi}{4}}.
\end{equation}

\noindent where \(\theta\) denotes the angles \(surv_1\) and \(surv_2\) make with the x-y plane in radians. The choice of \(\theta\) for \(surv_2\) presents an interesting problem. Since, the antennas are highly directive, picking a \(\theta\) close to 90 degrees implies that the target movement has to be very close to the antenna to be detected. In contrast, \(\theta\) of 0 degrees does not pick any movement in the x-z plane due to Doppler shift being 0. As a compromise, we choose a \(\theta\) of 45 degrees. In such a configuration, correlation signal at \(surv_2\) is given by \(s_2[n]\) and is formed by the superimposition of Doppler shifts in x-y and x-z planes, and can be represented by the following equation, i.e., 
\begin{equation}
{s_2[n] = \sum_{f_d \in f_{dxy}} Ax[n+\tau]e^{i2\pi f_{d}n} + \sum_{f_d \in f_{dxz}} Ax[n+\tau]e^{i2\pi f_{d}n}} ,
\end{equation}

\noindent where \(f_{dxy}\) and \(f_{dxz}\) contain a set of Doppler shifts in x-y and x-z planes, respectively. Similarly, correlation signal at \(surv_1\) is given by \(s_1[n]\) and is represented by the following equation, i.e., 

\begin{equation}
{s_1[n] = \sum_{f_d \in f_{dxy}} Ax[n+\tau]e^{i2\pi f_{d}n}}.
\end{equation}

\noindent Because of the geometry of surveillance antennas, extracted phase information from \(\phi_1[n]\) also leaks into \(\phi_2[n]\). In order for \(s_2[n]\) to only track movements in x-z plane, Vi-Wi calculates \(\phi_2^{'}[n]\) as
\begin{equation}
{\phi_2^{'}[n] = \phi_2[n] - \phi_1[n]},
\end{equation}

\noindent where \(\phi_1[n]\) and \(\phi_2[n]\) are phase signals calculated through Equation (8) from \(s_1[n]\) and \(s_2[n]\) respectively, and \(\phi_2^{'}[n]\) is a modified phase signal to account for information leakage from x-y plane to x-z plane.

\subsection{Feature Extraction}

\noindent After Vi-Wi extracts phases from surveillance signals, it proceeds by extracting features in following order: 

\begin{itemize}
\item Vi-Wi determines if there is any activity in the environment.
\item Vi-Wi classifies the dominant activity (the activity that results in the highest Doppler shift)
\item Vi-Wi extracts specific activity information. 
\end{itemize}

\noindent To illustrate these steps, let us consider the phase signals \(\phi_1[n]\) and \(\phi_2[n]\) associated with surveillance signals \(s_1[n]\) and \(s_2[n]\), against each activity. Fig. 3(a) shows that when a person inhales, her chest moves towards the device resulting in a positive Doppler shift; when she inhales, her chest moves away from the device causing a negative Doppler. Fig. 3(c) and 3(d) show variations in surveillance signals when the dominant activity is essential tremor. In such a scenario, both \(\phi_1\) and \(\phi_2\) alternate between positive and negative Doppler shifts. Finally, a human fall causes an aperiodic variation in \(\phi_2\) as shown in Fig. 3(f). 

\begin{figure}[!t]
\centering
\includegraphics[scale=0.34]{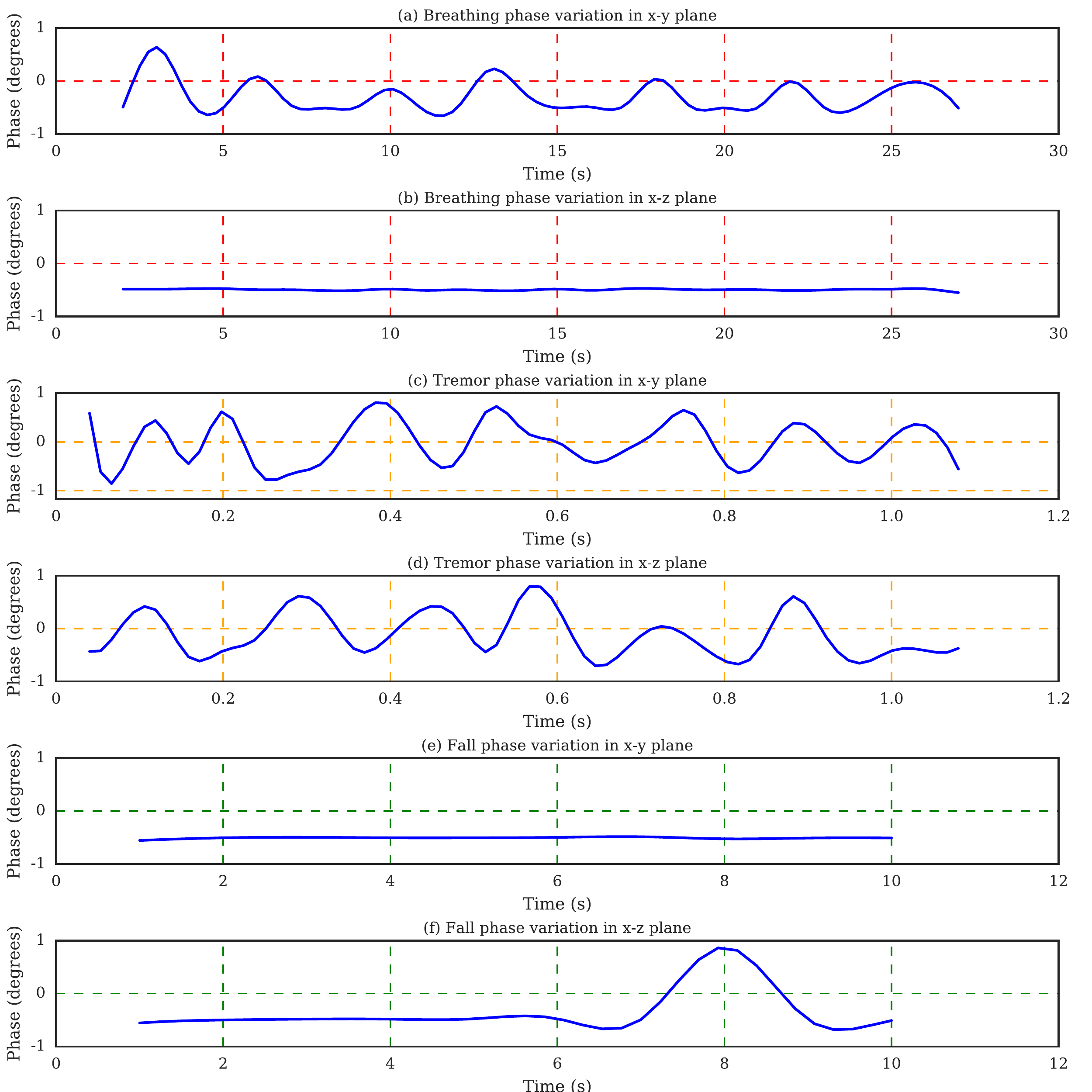}
\caption{Tracking \(\phi_1\) and \(\phi_2\) during breathing, tremor and human fall.}
\end{figure}

However, there may be instances when a user moves her limbs or makes some other dominant motion. To deal with such scenarios, Vi-Wi operates on a window of 20 seconds and determines if \(\phi_1\) and \(\phi_2\) are periodic. If both are aperiodic, we detect a random movement such as limb motion. However, if only \(\phi_2\) is aperiodic, this indicates a sudden movement in x-z plane and we detect a fall. If \(\phi_1\) is periodic and \(\phi_2\) is aperiodic, we recognize this as breathing motion in x-y plane only. If both are periodic, we detect tremor activity.

\noindent In the following section, we show how Vi-Wi determines the specifics of an activity once it has been classified.

\noindent\textit{Breathing Rate Measurement}

\noindent Fundamentally, the breathing rate can be extracted by taking fast Fourier transform (FFT) of the phase signal and subsequently picking the FFT peak. However, this does not provide an accurate estimate since the frequency resolution is quite small. Specifically, for a window size of 10 seconds, the frequency resolution is only 0.1\(Hz\) or approximately 6 breaths per second. In order to improve this resolution, we utilize a well known property that the dominant frequency of a signal can be accurately measured by doing regression on the phase of the complex time-domain signal [19]. Since, the phase of complex signal \(\phi_1[n]\) is linear, its slope corresponds to breathing rate estimate in \(Hertz\).

\noindent\textit{Tremor Classification}

\noindent Tremor measurement is similar to breathing rate measurement, but with two main variations. First, the dominant essential tremor frequency is in range 4-11 \(Hz\) as compared to 0.5-2 \(Hz\) in breathing. Second, the tremor measurements are taken across both x-y and x-z planes instead of just x-y plane.

\noindent Accurately determining the frequency in such resolution range is challenging. Moreover, the precise frequency estimate of essential tremor does not provide any added value to the health professionals. Therefore, we instead focus our experiments on classifying tremor as either low or high. The dominant frequency is measured using the same approach as in last passage, and then a classification decision is made based on whether the frequency is above or below a certain threshold. We note that if the frequency is too large or too small, we discard the activity as random motion.

\noindent\textit {Fall detection}

\noindent Fall detection is done by locating an instance of time when there is a spike in \(\phi_2[n]\) and \(\phi_1[n]\) is flat. We note that fall detection can be done continuously even when Vi-Wi is monitoring subject's breathing rate or essential tremor.

\section {Implementation}
\noindent The implementation settings are shown in Fig. 4 and explained below.

\subsection{Hardware}
\noindent The passive sensing system used in the experiments utilizes USRP B200 software defined radio with an omni-directional antenna as an access point. The access point transmits orthogonal frequency division multiplexed symbols at a data rate of 3 Mb/s with a code rate of \(\frac{1}{2}\) and quadrature phase shift keying modulation. With this configuration, the transmit power of the WiFi source is estimated to be around -60 dBm. At the receiver end, we have log-periodic directional antennas with 5dBi gain and 60 degree beam-width as shown in Fig. 4 on surveillance antenna 1 and 2. The signals received at these antennas are digitized through a Spartan 6 XC6SLX75 FPGA and 61.44 MS/s, 12 bit ADC. For non line-of-sight experiments (with obstacles), we use a 30 dB RF power amplifier with the transmit antenna.

\begin{figure}[!t]
\centering
\includegraphics[scale=0.47]{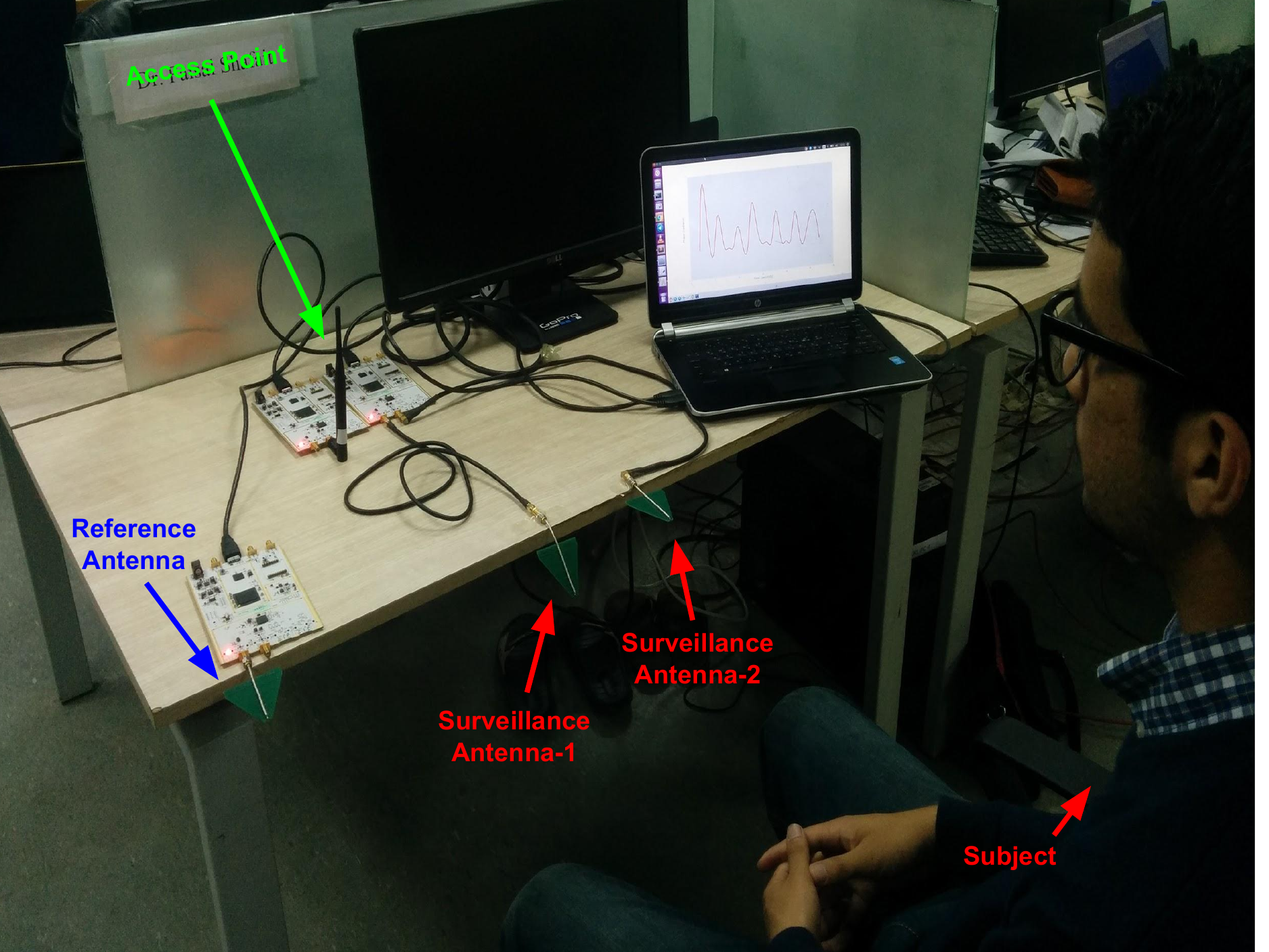}
\caption{Experiment Setup. The breathing rate of the subject is tracked and the activity is displayed on a personal computer.}
\end{figure}

\subsection{Software}
\noindent We implement Python blocks in GNU Radio for signal processing operations. The code runs in real-time and the display is updated on Python \textit{matplotlib} library every 2 seconds.

\subsection{Ground Truth}
\noindent In our experiments, we used a pulse oximeter to establish ground truth for breathing rate measurements. However, our study of essential tremor was limited as we could not find a reliable device to estimate tremor. Therefore, we developed a binary classification metric and determined if the tremor was high or low. In future studies, a more reliable ground truth could be established by using a reliable tremor monitoring device.

\section {Experimental Evaluation}
\begin{figure}[!t]
\centering
\includegraphics[scale=0.47]{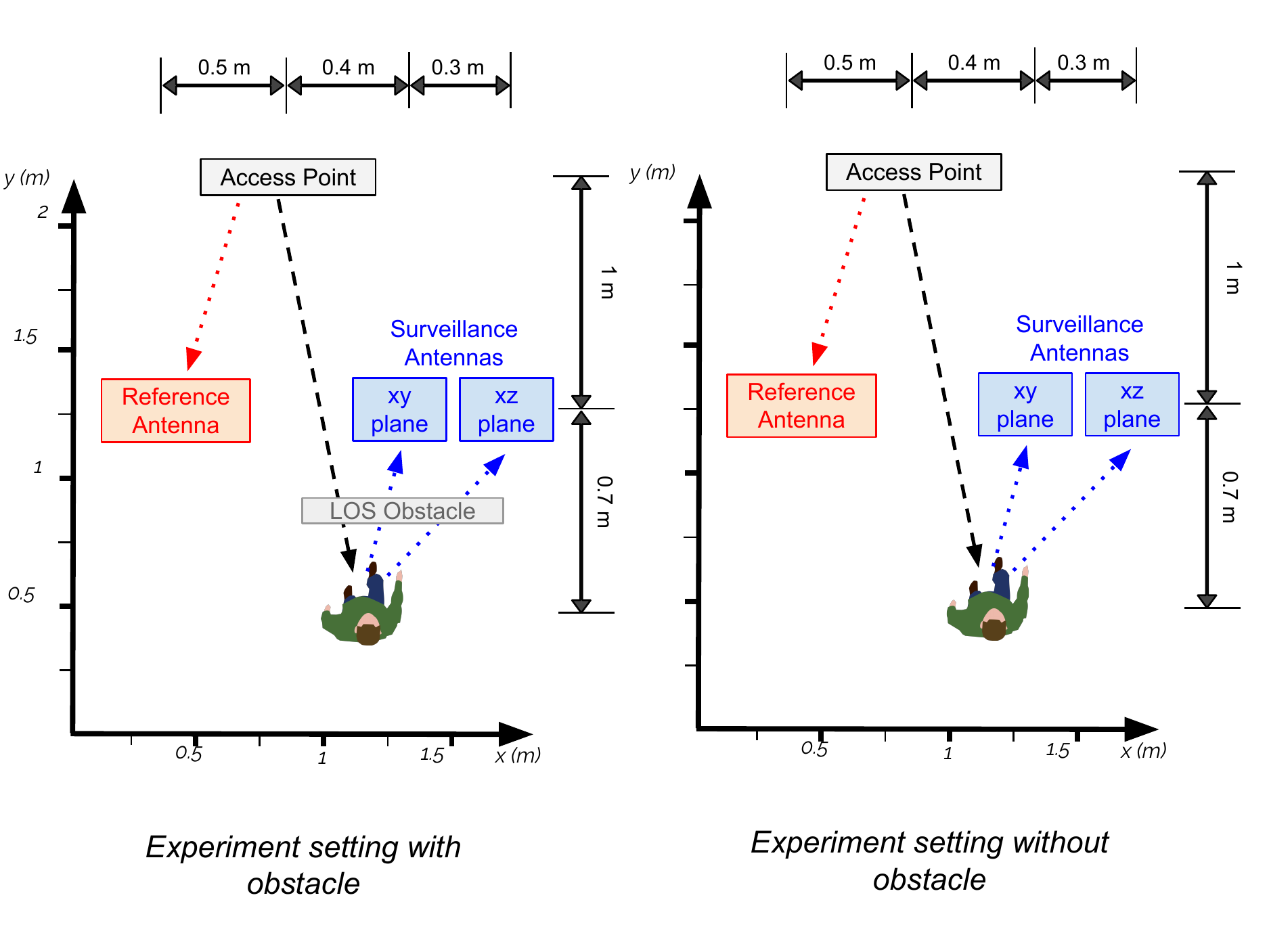}
\caption{Experiment settings for LOS and NLOS environments. There is a single access point, a reference antenna and two surveillance antennas. In NLOS setting, an additional metal block is placed to occlude the view of target.}
\end{figure}

\noindent To evaluate the performance of Vi-Wi, we performed experiments in a standard office environment with and without obstacles. A simple metal block was used as an obstacle, and it blocked all line of sight signals towards the receive antennas.

\begin{figure}[!t]
\centering
\includegraphics[scale=0.45]{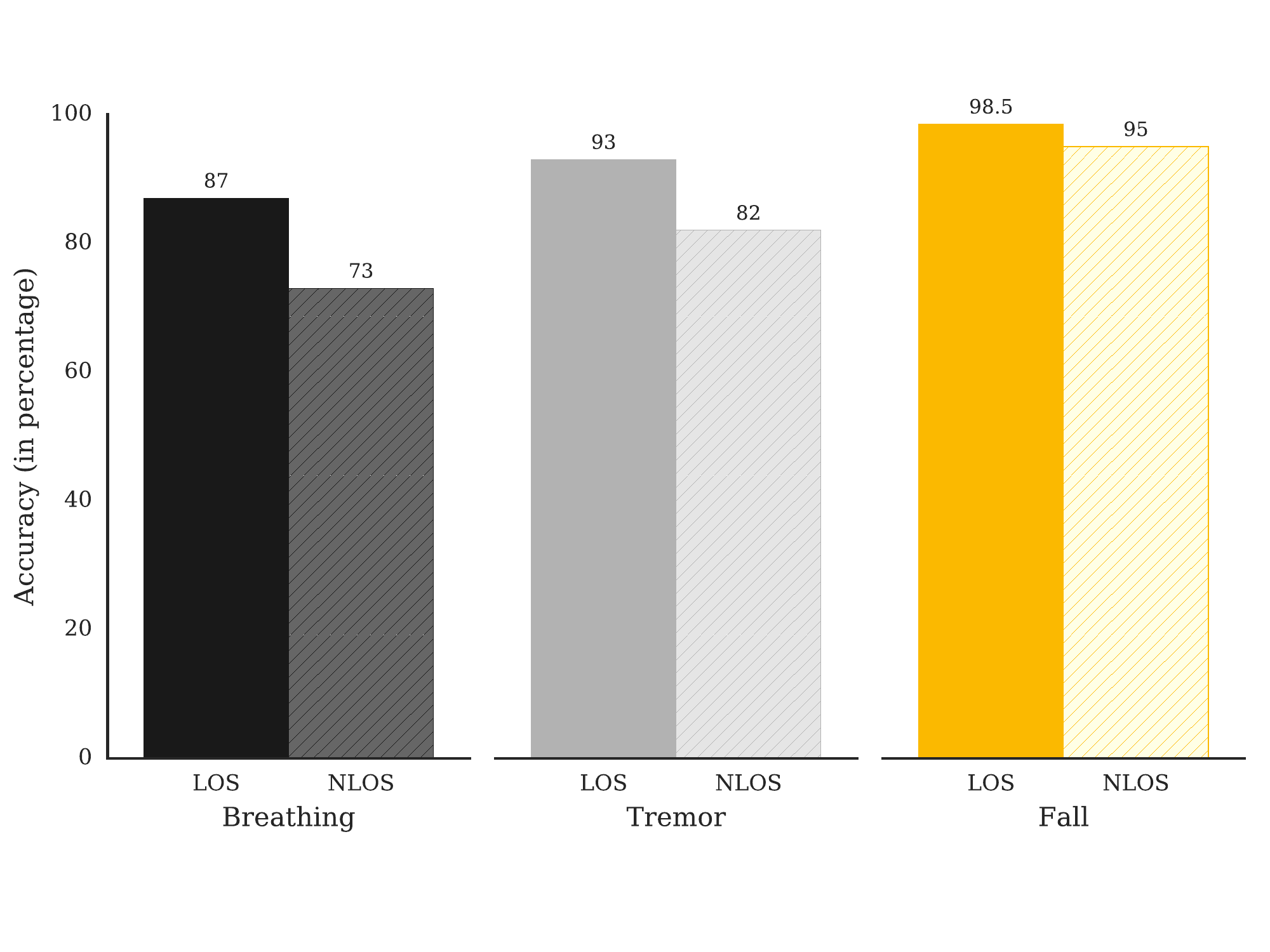}
\caption{Accuracy comparison of health activities. Breathing rate accuracy is determined by comparing actual breaths per minute to monitored breaths. Tremor classification accuracy is determined the ability to classify slow (4-7Hz) and fast (7-11Hz) tremor.   In our experiments, fall detection has the highest accuracy, while breathing has the least. Breathing measurement also undergoes the highest drop in accuracy from LOS to NLOS environment.}
\end{figure}

\noindent Fig. 6 shows Vi-Wi's accuracy for different elderly-care activities before and after an obstacle is placed in the line of sight of receiver. We observe that Vi-Wi is able to detect falls with the highest accuracy followed by tremor classification and breathing rate measurement. In the presence of an obstacle, the accuracy of all three activities considerably drops but we note that it can be improved by increasing the AP's transmit power or using antennas with higher gains. Here, we note several factors that may account for some of the inaccuracies in these small scale motion measurements. First, the wavelength of the WiFi signal is in order of centimeters and thus, slight geometry changes of human body may cause additional phase variations. Second, at each instance of time, a slightly different body part may be detected causing deviation in phase variations. Finally, these small scale motions are unlikely to be regular and a small change in orientation may cause a large phase deviation. 

\begin{figure}[!t]
\centering
\includegraphics[scale=0.58]{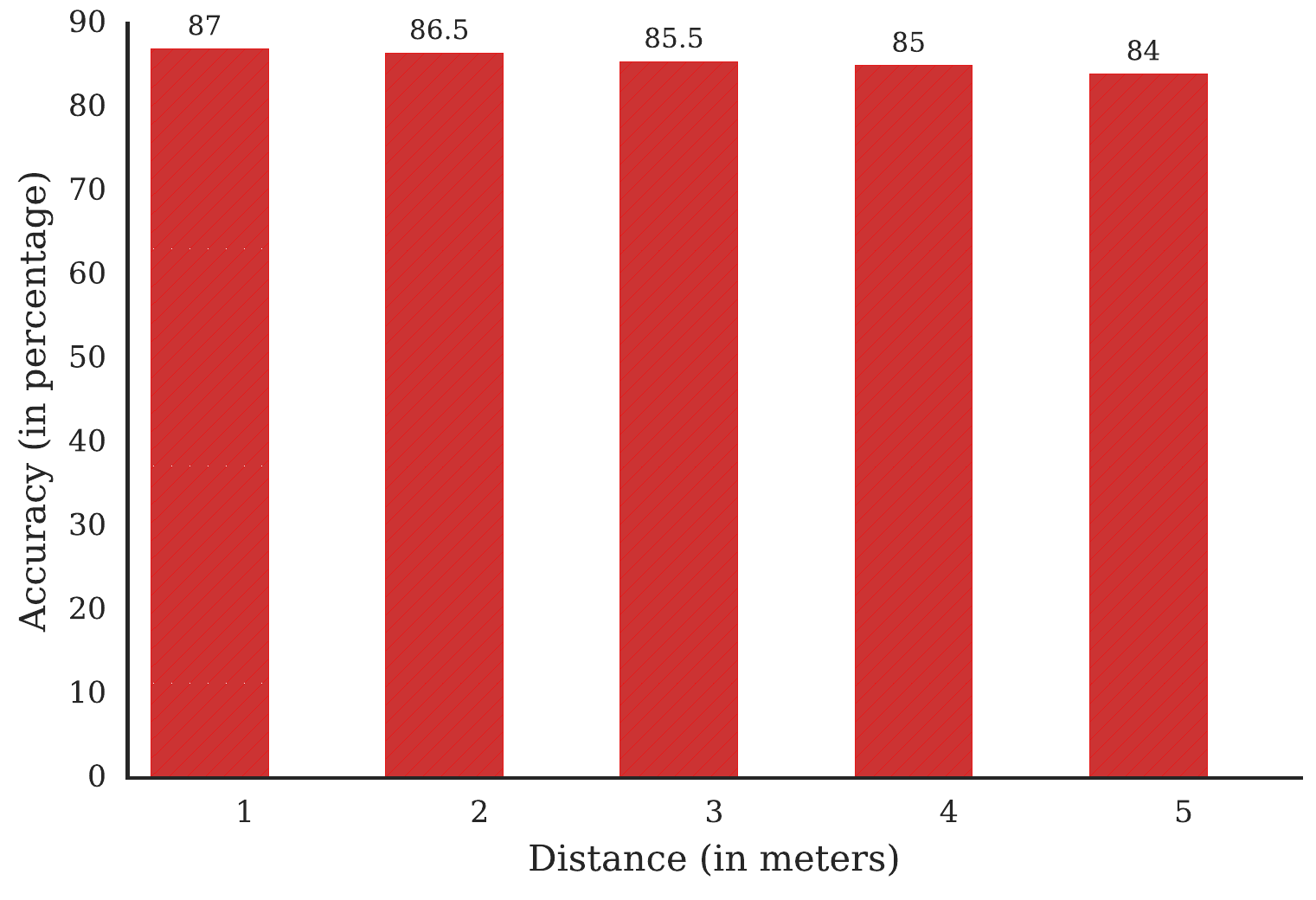}
\caption{Change in breathing rate accuracy with distance. Transmit power is set to 20 dBm. At a distance of 1 meter, the breathing rate accuracy is the highest at 87\%, and drops to 84\% when the target is 5 meters away.}
\end{figure}

Having fixed the transmit power at 20 dBm, we validated the performance of Vi-Wi as the distance of the target to device changed as shown in Fig. 7. At all the distances, the accuracy remained above 84\% while the highest accuracy (87\%) was achieved at the nearest distance of 1 meter. We then asked the participants to change their orientations and sit at slight angles facing the receive antennas. In particular, we asked the subjects to sit at three angles facing the antenna: \(30^0\), \(60^0\) and \(90^0\). At each distance, changing the sitting orientation did not have any significant effect on breathing rate accuracy. This is because during breathing, the chest expands in all directions and even though the subject is facing sideways, his chest movements can be detected. However, we noticed that changing the subject orientation caused tremor classification accuracy to drop  significantly. This is because unlike breathing, tremor motion is concentrated along fixed directions and as the orientation of hand changes, Doppler shift can no longer be monitored.

\begin{figure}[!t]
\centering
\includegraphics[scale=0.57]{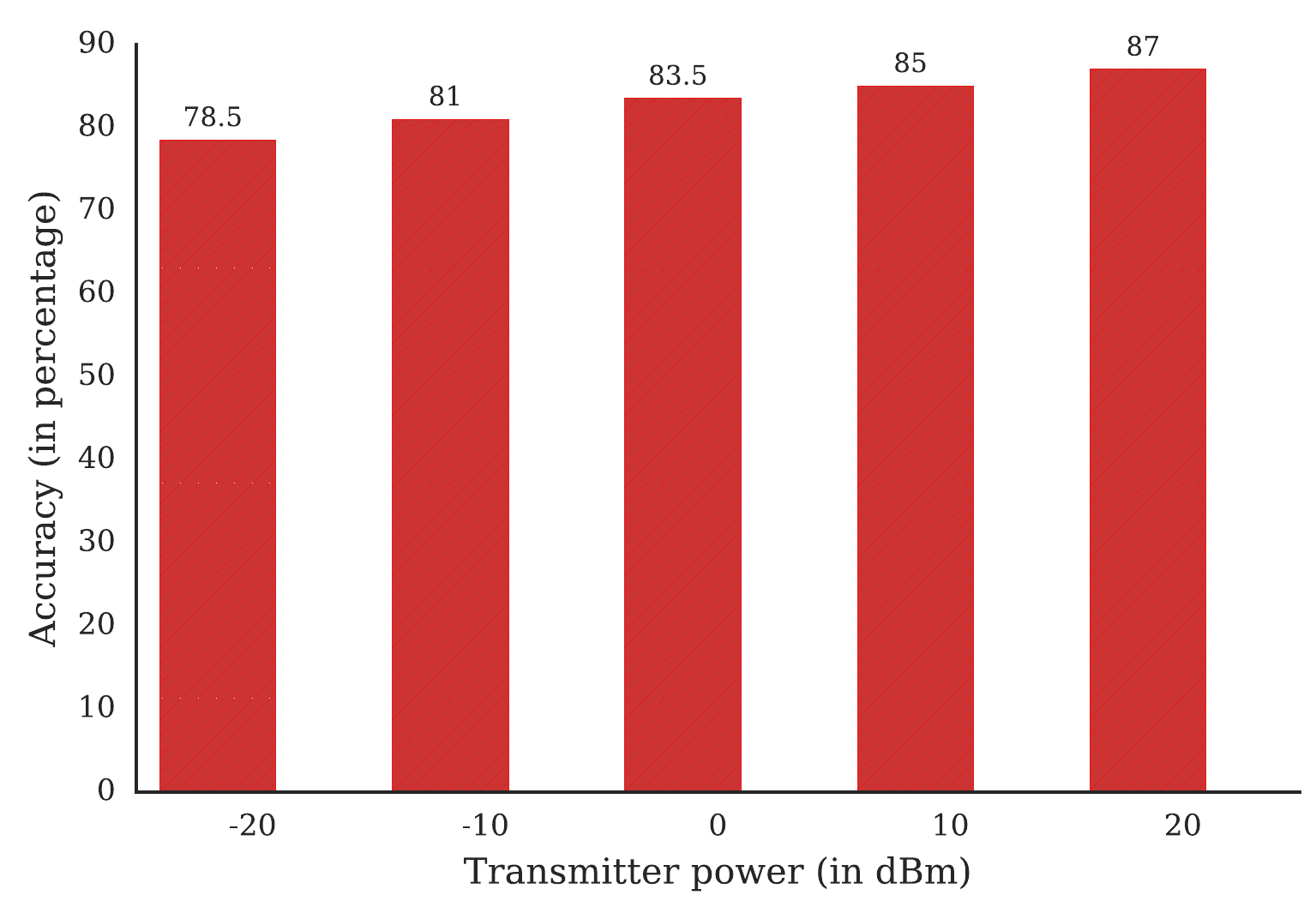}
\caption{Change in breathing rate accuracy with transmit power. Distance of the target is set to 1 meter. The accuracy improves by around 11\% when the transmit power is changed from -30 dBm (78.5\%) to 20 dBm (87\%).}
\end{figure}

Our final experiment was to analyze the effect of transmit power on system accuracy  as shown in Fig. 8. Our custom designed AP offered a maximum power of -10 dBm. By applying an external 30 dB RF power amplifier, Vi-Wi's breathing rate measurement accuracy improved by 7.4\%. There were similar improvements in the accuracies of tremor and fall activities with increase in transmit power. 

\subsection{Limitations}

In this section, we discuss some of the limitations of Vi-Wi:

\begin{itemize}
\item Vi-Wi assumes that there are no motion or background artefacts in the environment. While Vi-Wi can work reasonably well in presence of background noise, its performance is severely affected when there are multiple movements.

\item Vi-Wi may be prone to detecting non-human motion. For example, it may mistake fall of stick in the environment as a human fall.

\item Since, Vi-Wi requires a minimum signal-to-noise ratio (SNR), it works in a limited range of 5 meters. Beyond this distance, the accuracy of Vi-Wi drops signficantly.

\end{itemize}

\section{Conclusion and Future Directions}

\noindent The health applications using passive radar sensing have been largely untapped. The traditional methods to monitor breathing rate, tremor and falls are invasive and inconvenient. The modern methods using active radars require dedicated transmitters and high bandwidth. In contrast, Vi-Wi offers covert, license-free and convenient health monitoring. Future works can focus on various aspects of Vi-Wi. First, monitoring vital health activities in presence of motion and background artefacts is an interesting challenge. Second, additional health activities such as human gait can be investigated. Third, research can focus on using multiple access points for analyzing activities of more than one target. 

Beyond health, passive radar sensing offers a number of promising future directions. By utilizing the access points already available in buildings and intelligently assigning them to the targets of interest, one can achieve almost ubiquitous sensing.

\end{document}